\documentclass{cta-author}



{}
{}
{}
\usepackage{algorithm}
\usepackage{algorithmic}
\usepackage{subfig}
\usepackage{booktabs,caption}
\usepackage[flushleft]{threeparttable}
\usepackage[labelsep=quad]{caption}
\begin{document}

\supertitle{This paper is a preprint of a paper accepted by IET Intelligent Transport Systems and is subject to Institution of Engineering and Technology Copyright. When the final version is published, the copy of record will be available at the IET Digital Library.}

\title{Hierarchical Reinforcement Learning for Self-Driving Decision-Making without Reliance on Labeled Driving Data}

\author{\au{Jingliang Duan$^{1}$}, \au{Shengbo Eben Li$^{1\corr}$}, Yang Guan$^{1}$, Qi Sun$^{1}$, \au{Bo Cheng$^{1}$}}

\address{\add{1}{School of Vehicle and Mobility, Tsinghua University, Beijing, 100084, China}
\email{lisb04@gmail.com}}

\begin{abstract}
Decision making for self-driving cars is usually tackled by manually encoding rules from drivers' behaviors or imitating drivers' manipulation using supervised learning techniques. Both of them rely on mass driving data to cover all possible driving scenarios. This paper presents a hierarchical reinforcement learning method for decision making of self-driving cars, which does not depend on a large amount of labeled driving data. This method comprehensively considers both high-level maneuver selection and low-level motion control in both lateral and longitudinal directions. We firstly decompose the driving tasks into three maneuvers, including driving in lane, right lane change and left lane change, and learn the sub-policy for each maneuver. Then, a master policy is learned to choose the maneuver policy to be executed in the current state. All policies including master policy and maneuver policies are represented by fully-connected neural networks and trained by using asynchronous parallel reinforcement learners (APRL), which builds a mapping from the sensory outputs to driving decisions. Different state spaces and reward functions are designed for each maneuver. We apply this method to a highway driving scenario, which demonstrates that it can realize smooth and safe decision making for self-driving cars.
\end{abstract}

\maketitle

\section{Introduction}\label{sec:intro}

Autonomous driving is a promising technology to enhance road safety, ease the road congestion, decrease fuel consumption and free the human drivers. In the architecture of the self-driving car, decision making is a key component to realize autonomy. To date, most of decision making algorithms can be categorized into two major paradigms: rule-based method and imitation-based method. The former is usually tackled by manually encoding rules from driver behaviors, whereas the latter is committed to imitate drivers' manipulation using supervised learning techniques. To achieve safe decision making under different traffic conditions, both of them rely on mass naturalistic driving data to cover all possible driving scenarios for the purpose of testing or training \cite{paden2016survey,bojarski2017NVIDA}. 

The rule-based method has been widely investigated during the last two decades, which is typically hierarchically structured into maneuver selection and path planning \cite{katrakazas2015rule-based}. Montemerlo et al. (2008) adopted finite state machines (FSM) as a mechanism to select maneuver for Junior in DARPA \cite{montemerlo2008junior}. The FSM possessed 13 states, including lane keeping, parking lot navigation, etc., which was then used to switch between different driving states under the guidance of manually modeling behavior rules. Furda et al. (2011) combined FSM with multi-criteria decision making for driving maneuver execution \cite{furda2011enabling}. The most appropriate maneuver was selected by considering the sensor measurements, vehicular communications, traffic rules and a hierarchy of objectives during driving with manually specified weights. Glaser et al. (2010) presented a vehicle path-planning algorithm that adapted to traffic on a lane-structured infrastructure such as highways \cite{glaser2010maneuver}. They firstly defined several feasible trajectories expressed by polynomial with respect to the environment, aiming at minimizing the risk of collision. Then they evaluated these trajectories using additional performance indicators such as travel time, traffic rules, consumption and comfort to output one optimal trajectory in the following seconds. Kala and Warwick (2013) presented a planning algorithm to make behavior choice according to obstacle motion in relatively unstructured road environment \cite{kala2013motion}. They modeled the various maneuvers in an empirical formula and adopted one of them to maximize the separation from other vehicles. Despite the popularity of rule-based methods, manually encoding rules can introduce a high burden for engineers, who need to anticipate what is important for driving and foresee all the necessary rules for safe driving. Therefore, this method is not always feasible due to the highly dynamic, stochastic and sophisticated nature of the traffic environment and the enormous number of driving scenarios \cite{duan2017driver,hou2019drivers,li2019deep}. 

The imitation-based method is an effective remedy to eliminate engineers' burden. This is because that the driving decision model can be learned directly by mimicking drivers' manipulation using supervised learning techniques, and no hand-crafted rules are needed \cite{lecun2015deep}. This idea dates back to the late 1980s when Pomerleau built the first end-to-end decision making system for NAVLAB \cite{pomerleau1989alvinn}. This system used a fully-connected network which took images and a laser range finder as input and steering angles as output. Inspired by this work, Lecun et al. (2004) trained a convolutional neural network (CNN) to make a remote truck successfully drive on unknown open terrain while avoiding any obstacles such as rocks, trees, ditches, and ponds \cite{lecun2004dave}. The CNN constructed a direct mapping from the pixels of the video cameras to two values which were interpreted directly as a steering wheel angle. With the development of deep learning, Chen et al. (2015) trained a deep CNN using data recorded from 12 hours of human driving in a video game \cite{chen2015deepdriving}. This CNN mapped an input image to a small number of key perception indicators that directly related to the affordance of a road/traffic state for driving. These methods had not been applied in practice until Bojarski et al. (2016) built an end-to-end learning system which could successfully perform lane keeping in traffic on local roads with or without lane markings and on highways. A deep CNN called PilotNet was trained using highway road images from a single front-facing camera only paired with the steering angles generated by a human driving a data-collection car \cite{bojarski2017NVIDA}. A shortcoming of aforementioned works is that their capability is derived from large amounts of hand-labeled training data, which come from cars driven by humans while the system records the images and steering angles. In many real applications, the required naturalistic driving data can be too huge due to various driving scenarios, large amounts of participants and complex traffic. Besides, different human drivers may make completely different decisions in the same situation, which results in an ill-posed problem that is confusing when training a regressor.

To eliminate the demand for labeled driving data, some researches have made attempts to implement reinforcement learning (RL) on decision making for self-driving cars. The goal of RL is to learn policies for sequential decision problems directly using samples from the emulator or experiment, by optimizing a cumulative future reward signal \cite{sutton2018reinforcement}. Compared with imitation-based method, RL is a self-learning algorithm which allows the self-driving car to optimize its driving performance by trial-and-error without reliance on manually designed rules and human driving data \cite{mnih2015DQN, silver2016Alphago,silver2017mastering}. Lillicrap et al. (2015) presented a deep deterministic policy gradient model to successfully learn deterministic control policies in TORCS simulator with an RGB images of the current frame as inputs \cite{lillicrap2015DPG}. This algorithm maintained a parameterized policy network which specified the current policy by deterministically mapping states to a specific action such as acceleration. Mnih et al. (2016) proposed an asynchronous advantage actor-critic algorithm (A3C) which enabled the vehicle to learn a stochastic policy in TORCS \cite{mnih2016A3C}. The A3C used parallel actor-learners to update a shared model which could stabilize the learning process without experience replay. However, decision making for self-driving cars remains a challenge for RL, because it requires long decision sequences or complex policies \cite{daniel2016probabilistic}. Specifically, decision making for self-driving cars includes higher-level maneuver selection (such as lane change, driving in lane and etc.) and low-level motion control in both lateral and longitudinal directions \cite{wei2014behavioral}. However, current works of RL only focus on low-level motion control layer. It is hard to solve driving tasks that require many sequential decisions or complex solutions without considering the high-lever maneuver \cite{paden2016survey}.

The main contribution of this paper is to propose a hierarchical RL method for decision making of self-driving cars, which does not depend on a large amount of labeled driving data. This method comprehensively considers both high-level maneuver selection and low-level motion control in both lateral and longitudinal directions. We firstly decompose the driving tasks in to three maneuvers, including driving in lane, right lane change and left lane change, and learn the sub-policy for each maneuver. Different state spaces and reward functions are designed for each maneuver. Then, a master policy is learned to choose the maneuver policy to be executed in the current state. All policies including master policy and maneuver policies are represented by fully-connected neural networks and trained by using asynchronous parallel RL algorithm, which builds a mapping from the sensory outputs to driving decisions. 

The rest of this paper is organized as follows: Section \ref{sec.methodology} proposes the hierarchical RL method for decision making of self-driving cars. Section \ref{sec:experiment} introduces the design of experiment, state space and return function. Section \ref{sec:results} discusses the training results for each maneuver and driving task. Section \ref{sec:conclusions} concludes this paper.

\section{Methodology}
\label{sec.methodology}

\subsection{Preliminaries}
\label{sec.preliminaries}

We model the sequential decision making problem for self-driving as a Markov decision process (MDP) which comprises: a state space $\mathcal{S}$, an action space $\mathcal{A}$, transition dynamics $p(s_{t+1}|s_t,a_t)$ and reward function $r(s_t,a_t): \mathcal{S} \times \mathcal{A} \rightarrow \mathbb{R}$ \cite{sutton2018reinforcement}. At each time step $\emph{t}$, the vehicle observes a state $s_t$, takes an action $a_t$, receives a scalar reward $r$ and then reaches the next state $s_{t+1}$. A stochastic policy $\pi(a|s): \mathcal{S} \rightarrow \mathcal{P}(\mathcal{A})$ is used to select actions, where  $\mathcal{P}(\mathcal{A})$ is the set of probability measures on measures on $\mathcal{A}$. The reward $r$ is often designed by human experts, which is usually assigned a lower value when the vehicle enters a bad state such as collision, and a higher value for a normal state such as maintaining a safe distance from the preceding car. This process continues until the vehicle reaches a terminal state, such as breaking traffic regulations or arriving at the destination. 

The return $R_t = \sum_{k=0}^{\infty}\gamma^kr_{t+k}$ is the sum of discounted future reward  from time step $t$, where $\gamma\in(0,1]$ is a discount factor that trades off the importance of immediate and future rewards. The goal of RL is to maximize the expected return from each state $s_t$. In general, the expected return for following policy $\pi$ from the state $s$ is defined as $V_{\pi}(s)=E[R_t|s_t=s]$, often referred to as a value function. Then the RL problem can now be expressed as finding a policy $\pi$ to maximize the associated value function $V_{\pi}(s)$ in each state.

This study employs deep neural network (NN) to approximate both the value function $V_{\pi}(s; w)$ and stochastic policy $\pi_{\theta}(s)$ in an actor-critic (AC) architecture, where $\theta$ and $w$ are the parameters of these two NNs. The AC architecture consists of two structures: 1) the  actor, and 2) the critic \cite{sutton2000actor-critic,degris2012actor-critic}. The actor corresponds to the policy network  $\pi_{\theta}(s)$, mapping states to actions in a probabilistic manner. And the critic corresponds to the value network $V_{\pi}(s; w)$, mapping states to the expected cumulative future reward. Thus, the critic addresses the problem of prediction, whereas the actor is concerned with choosing control. These problems are separable, but are solved by iteratively solving the Bellman optimality equation based on generalized policy iteration framework \cite{bhatnagar2009natural}.

\subsection{Hierarchical parallel reinforcement learning algorithm}

\subsubsection{Hierarchical RL architecture}

\begin{figure*}[tb]
\centering{\includegraphics[width=0.75\textwidth]{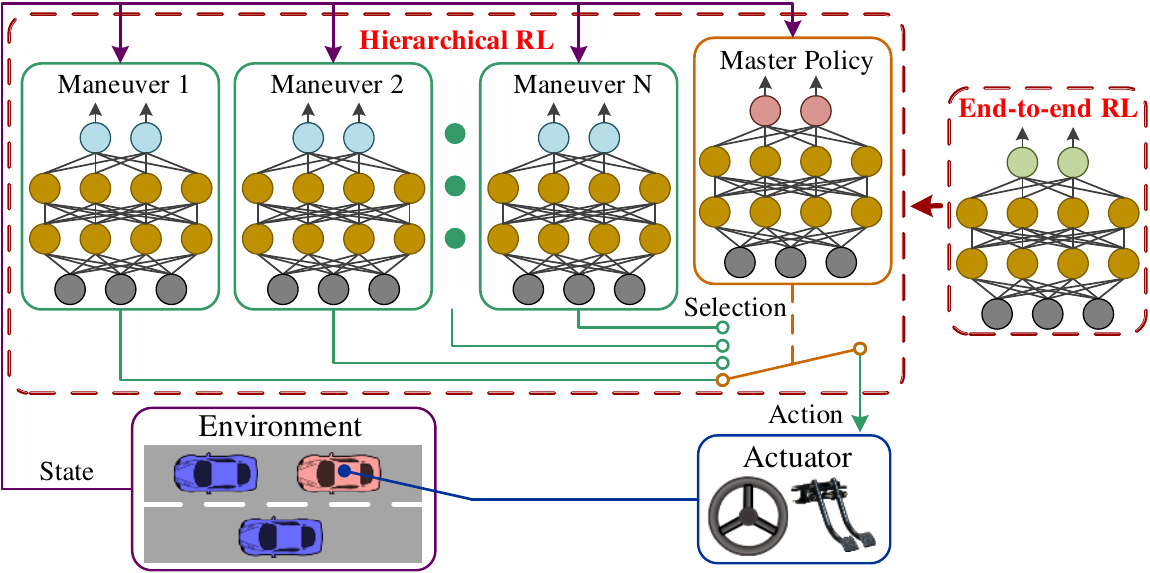}}
\caption{{\normalsize \textit{Hierarchical RL for self-driving decision-making}}\label{f:archi}}
\end{figure*}

Previous related RL studies typically adopted an end-to-end network to make decisions for autonomous vehicles. To complete a driving task consisting of various driving behaviors, complicated reward functions need to be designed by human experts. Inspired by the option framework \cite{Sutton1999option,gopalan2017planning}, we propose a hierarchical RL (H-RL) architecture to eliminate the burden of engineers by decomposing the driving decision making processes into two levels. The key insight behind our new method is illustrated in Fig. \ref{f:archi}. The H-RL decision making framework consists of two parts: 1) the high-level maneuver selection and 2) low-level motion control. In the maneuver selection level, a master policy is used to choose the maneuver to be executed in the current state. In the motion control level, the corresponding maneuver policy would be activated and output front wheel angle and acceleration commands to the actuators. In the example shown in Fig. \ref{f:archi}, the master policy chooses maneuver 1 as the current maneuver, then the corresponding maneuver policy would be activated and output front wheel angle and acceleration commands to the actuators. The combination of multiple maneuvers can constitute diverse driving tasks, which means that the trained maneuver policy can also by applied to other driving tasks. Therefore, hierarchical architecture also has better transfer ability than end-to-end RL in the field of driving decision.

To learn each policy of the H-RL, we should first decompose the driving task into several driving maneuvers, such as driving in lane, left/right lane change, etc. It should be noted that a driving maneuvers may include many similar driving behaviors. For example, the driving-in-lane maneuver is a combination of many similar behaviors including lane keeping, car following and free driving. Then, we can learn the sub-policy, which is also called maneuver policy, for each maneuver driven by independent sub-goals. The reward function for sub-policy learning can be easily designed by considering only the corresponding maneuver instead of the whole driving task. Then we learn a master policy to activate certain maneuver policy according to the driving task. Although we need to consider the entire driving task while training the master policy, the associated reward function can also be very simple because you do not have to worry about how to control the actuators to achieve each maneuver. In addition, if the input to each policy is the whole sensory information, the learning algorithm must determine which parts of the information are relevant. Hence, we propose to design different meaningful indicators as the state representation for different policies. 

Furthermore, the motion of the vehicle is controlled jointly by the lateral and longitudinal actuators, and the two types of actuators are relatively independent. This means that the state and reward of the lateral and longitudinal controller can also be designed separately. Therefore, each maneuver in this study contains a steer policy network (SP-Net) and an accelerate policy network (AP-Net), which carry out lateral and longitudinal control respectively. Noted that the vehicle dynamics used in this paper still consider the coupling between the lateral and longitudinal controllers. The corresponding value networks of these two control policy networks are called SV-Net and AV-Net respectively. In other word, SP-Net and SV-Net are concerned with lateral control, while AP-Net and AV-Net are responsible for longitudinal control. On the other hand, the master policy only contains one maneuver policy network and one maneuver value network. In summary, if a maneuver is selected by the master policy, the relevant SP-Net and AP-Net would work simultaneously to apply control. 

\subsubsection{Parallel training algorithm}
\begin{figure}[b]
\centering{\includegraphics[width=0.48\textwidth]{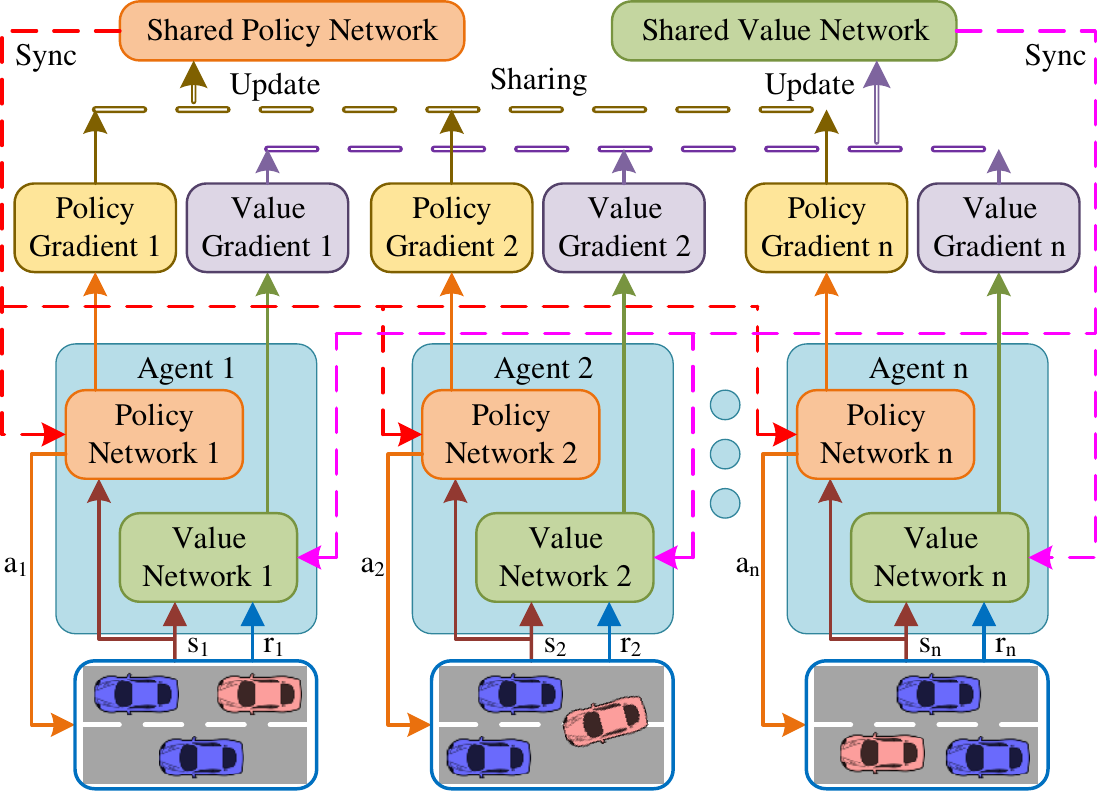}}
\caption{{\normalsize \textit{Policy training using APRL}} \label{f:prl}}
\end{figure}

Inspired by the A3C algorithm, we use asynchronous parallel car-learners to train the policy $\pi(a|s;\theta)$ and estimate the state value $V(s; w)$, as shown in Fig. \ref{f:prl} \cite{mnih2016A3C}. Each car-learner contains its own policy networks and a value networks, and makes decision according to their policy outputs. The multiple car-learners interact with different parts of the environment in parallel. Meanwhile, each car-learner computes gradients with respect to the parameters of the value network and the policy network at each step. Then, the average gradients are applied to update the shared policy network and shared value network at each step. These car-learners synchronize their local network parameters from the shared networks before they make new decisions. We refer to this training algorithm as asynchronous parallel reinforcement learning (APRL).

For each car-learner, the parameters $w$ of value network $V(s; w)$ are tuned by iteratively minimizing a sequence of loss functions, where the loss function at time step $t$ is defined as:
\begin{equation}
L_t(w) = (R_t-V(s_t; w))^2
\end{equation}
where $R_t-V(s_t; w)$ is usually called temporal-difference (TD) error. The expected accumulated return $R_t$ is estimated in the forward view using N-step return instead of full return:
\begin{equation}
R_t = \sum_{k=0}^{N-1}\gamma^kr_{t+k} + \gamma^nV(s_{t+n}; w)
\end{equation}

This means that the update gradients for value and policy networks at time $t$ are estimated after $n$ actions. The specific gradient update for the parameters $w$ of $V(s; w)$ after state $s_t$ is:
\begin{equation}
dw = (R_t-V_{w}(s_t))\nabla_{w} V_{w}(s_t)
\end{equation}

The policy gradient of the actor-critic method is $(R_t-V_{w_t}(s_t))\nabla_{\theta}\log\pi(a_t|s_t;\theta)$ \cite{sutton2000actor-critic}. Besides, the entropy of the policy $\sum_a(-\pi_{\theta}(a_t|s_t)\log\pi_{\theta}(a_t|s_t))$ is added to the objective function to regularize the policy towards larger entropy, which promotes exploration by discouraging premature convergence to suboptimal policies. In summary, the total gradient consists of the policy gradient term and the entropy regularization term. Then, the specific gradient update for the parameters $\theta$ of the policy network after state $s_t$ is:
\begin{multline}
d\theta = (R_t-V_{w}(s_t))\nabla_{\theta} \log\pi_{\theta}(a_t|s_t) + \\\beta\nabla_{\theta}\sum_a(-\pi_{\theta}(a_t|s_t)\log\pi_{\theta}(a_t|s_t))
\end{multline}
where $\beta$ is the hyper-parameter that trades off the importance of different loss components. The pseudocode of APRL is presented in Algorithm \ref{alg1}. The parameters of the shared value and policy networks are represented by $w$ and $\theta$, while those of the value and policy networks for specific car-learners are denoted by $w'$ and $\theta'$. 

\begin{algorithm}[!htb]
  \captionsetup{font={footnotesize}}
  \caption{APRL algorithm}
  \label{alg1}
  \footnotesize
  \begin{algorithmic}
    \STATE Initialize $w$, $\theta$, $w'$, $\theta'$, step counter $t \gets 0$ and state $s_0 \in S$
    \REPEAT
	\FOR{each car-learner}
    \STATE Synchronize car-learner-specific parameters $w' = w$ and $\theta' = \theta$
	\STATE Perform $a_t$ according to policy $\pi(a_t|s_t; \theta')$ and store action $a_t$ 
	\STATE Receive and store reward $r_t$ and the next state $s_{t+1}$
	\STATE $\tau \gets t - n + 1$ 
	\IF {$\tau \ge 0$} 
	\STATE $R \gets \sum_{k=\tau}^{min(\tau + N - 1, t_{max})}\gamma^{k-\tau}r_k + \gamma^{k-\tau+1} V(s_{t+1}; w')$
	\STATE  Calculate gradients wrt $w'$: 
	\STATE \qquad $dw' = (R-V_{w'}(s_{\tau}))\nabla_{w'} V_{w'}(s_{\tau})$ 
	\STATE Calculate gradients wrt $\theta'$: 
    \STATE \qquad $ d\theta' =  (R-V_{w'}(s_{\tau}))\nabla_{\theta'} \log\pi_{\theta'}(a_{\tau}|s_{\tau}) + $ 
    \STATE \qquad \qquad \qquad \qquad  $\beta\nabla_{\theta'}\sum_{a}(-\pi_{\theta'}(a_{\tau}|s_{\tau})\log\pi_{\theta'}(a_{\tau}|s_{\tau}))$ 
	\ENDIF
	\ENDFOR
	\STATE $t \gets t+1$
	\STATE Update $w$ using $\overline{dw'}$ 
	\STATE Update $\theta$ using $\overline{d\theta'}$
	\UNTIL Convergence  
  \end{algorithmic}
\end{algorithm}

\section{Experiment design}\label{sec:experiment}
\subsection{Driving task description}
This paper focuses on two-lane highway driving. The inner (or left) lane is a high-speed lane with the speed limit from 100 to 120 $\rm km \slash h $, and the outer (or right) lane is a low-speed lane with the speed limit from 60 to 100 $\rm km \slash h $ \cite{liao2016detection}. The self-driving car would be initialized at a random position of these two lanes. The destination is also a random set on the inner lane or outer lane with a random distance between 500 m to 1000 m. The self-driving car needs to learn how to reach the destination as fast as possible without breaking traffic rules or colliding with other cars. If the initial position and destination are both on the low-speed lane, the car has to change lane at least twice during the driving task. At first, the car needs to change to the high-speed lane and keeps driving for a while, and then switches back to the low-speed lane when it approaches the destination. Continuous surrounding traffic is presented in both lanes to imitate real traffic situations, and is randomly initialized at the beginning of the task. The continuous traffic flow means that there are always other cars driving around the ego vehicle. The movement of other cars is controlled by the car-following model presented by Toledo \cite{toledo2007traffic}. The dynamics of the self-driving car is approximated by a dynamic bicycle model \cite{li2017driver}. In addition, the system frequency is 40Hz.
\subsection{Action space}
Previous RL studies on autonomous vehicle decision making usually take the front wheel angle and acceleration as the policy outputs \cite{lillicrap2015DPG,mnih2016A3C}. This is unreasonable because the physical limits of the actuators are not considered. To prevent large discontinuity of the control commands, the outputs of maneuver policy SP-Net and AP-Net are front wheel angle increment and acceleration increment respectively, i.e.,
\begin{equation}
A_{\text{sub}} =
\left\{
\begin{tabular}{lll}
 $-0.8:0.2:0.8$ & $^{\circ}/\rm{Hz}$ & SP-Net\\
 $-0.2:0.05:0.2$ & $(\rm m \slash s^2)/\rm{Hz} $ & AP-Net\\
\end{tabular}
\right.
\end{equation}
The outputs of SP-Net and AP-Net are designed with reference to the previous researches \cite{attia2012coupled,katriniok2013optimal,gray2012semi}. On the other hand, the highway driving task in this study is decomposed into three maneuvers: driving in lane, left lane change and right lane change. As mentioned above, the driving-in-lane maneuver is a combination of many behaviors including lane keeping, car following and free driving. These three maneuvers are proved to account for more than 98\% of the highway driving \cite{li2017estimation}. Therefore the action space the master policy is defined as follows:
\begin{equation}
A_{\text{master}}=
\left\{
\begin{tabular}{ll}
 0 & driving in lane\\
 1 & left lane change\\
 2 & right lane change\\
\end{tabular}
\right.
\end{equation}

\subsection{State representation and reward function}
\begin{figure}[tb]
\centering{\includegraphics[width=0.3\textwidth]{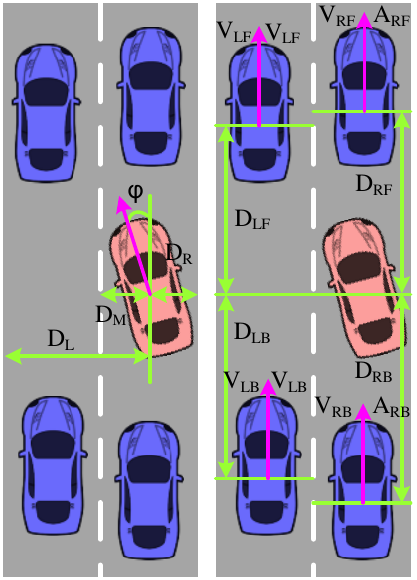}}
\caption{{\normalsize \textit{Illustration of state representations}}\label{f:state}}
\end{figure}

We propose four types of states to represent driving scenario and task: states related to the host car, states related to the road, states related to other cars and states related to the destination. It should be noted that we only involve four surrounding cars in this problem: a preceding car and a following car in each lane. In total, we propose 26 indicators to represent the driving state, some of which are illustrated in Fig. \ref{f:state}. A complete list of the states is given in Table \ref{tab.allstate}. Each state is added with a noise from uniform distribution before it is observed by the car-learner. We select a part of the states from the list as inputs to the relevant policy networks and value networks. The state space and reward function of each network are varied and depend on the type of maneuver and action.

\begin{table*}[!b]
\begin{threeparttable}
\processtable{{\normalsize State representation}\label{tab.allstate}}
{\begin{tabular*}{18cm}{p{1.5cm}<{\centering}p{1cm}<{\centering}p{1cm}<{\centering}p{1.5cm}<{\centering}p{1.25cm}<{\centering}p{1.25cm}<{\centering}p{1.25cm}<{\centering}p{1.25cm}<{\centering}p{1.5cm}<{\centering}p{1.5cm}<{\centering}p{1.5cm}<{\centering}}
\toprule
type & note  & unit & max noise\tnote{a}& \multicolumn{4}{c}{maneuver 0} & maneuver 1 & maneuver 2 &  master task\\
\cmidrule{5-8}
&&&& \multicolumn{2}{c}{left lane} & \multicolumn{2}{c}{right lane} &&& \\
\cmidrule(lr){5-6} \cmidrule(lr){7-8} 
&&&& SP-Net& AP-Net &  SP-Net & AP-Net &&& \\
\midrule
host car\tnote{b}& $\rm \delta$ &   $\rm ^{\circ}$ & 0.2 &\checkmark & &\checkmark&&\checkmark&\checkmark&\checkmark\\
& V &  $\rm km \slash h $ & 0.5 &\checkmark &\checkmark &\checkmark &\checkmark &\checkmark&\checkmark&\checkmark\\
& A &  $\rm m \slash s^2 $ & 0.1 &\checkmark &\checkmark &\checkmark &\checkmark &\checkmark&\checkmark&\checkmark\\
\midrule
 road\tnote{c}& $\rm \phi $ & \rm $^{\circ}$  & 0.5 &\checkmark&  &\checkmark&&\checkmark&\checkmark&\checkmark\\
 &$\rm D_L$ & m & 0.05 &\checkmark&&&&&\checkmark&\checkmark\\
 &$\rm D_M$ & m & 0.05 &\checkmark&&\checkmark&&\checkmark&\checkmark&\checkmark\\
 &$\rm D_R$ & m & 0.05&&&\checkmark&&\checkmark&&\checkmark\\
 &$\rm L_C$ & - & -&&&&&&&\checkmark\\
 &$\rm V_{LU}$  &  $\rm km \slash h $ & -&&\checkmark&&&\checkmark&& \checkmark\\
 &$\rm V_{LL}$ &  $\rm km \slash h $ &  -&&\checkmark&&&\checkmark&&\checkmark\\
 &$\rm V_{RU}$  &  $\rm km \slash h $  & -&&&&\checkmark&&\checkmark&\checkmark\\
 &$\rm V_{RL}$ &  $\rm km \slash h $ &  - &&&&\checkmark&&\checkmark&\checkmark\\
\midrule
  other cars&$\rm D_{LF}$ &  m & 1&&\checkmark&&&\checkmark&\checkmark&\checkmark\\
 &$\rm V_{LF}$ &   $\rm km \slash h $ & 1&&\checkmark&&&\checkmark&\checkmark&\checkmark\\
 &$\rm A_{LF}$ &    $\rm m \slash s^2 $ & 0.3&&\checkmark&&&\checkmark&\checkmark&\checkmark\\
 &$\rm D_{LB}$ &    m  & 1&&&&&\checkmark&&\checkmark\\
 &$\rm V_{LB}$ &    $\rm km \slash h $& 1 &&&&&\checkmark&&\checkmark\\
 &$\rm A_{LB}$ &    $\rm m \slash s^2 $ & 0.3&&&&&\checkmark&&\checkmark\\
 &$\rm D_{RF}$ &    m & 1&&&&\checkmark&\checkmark&\checkmark&\checkmark\\
 &$\rm \rm V_{RF}$    & $\rm km \slash h $&  1&&&&\checkmark&\checkmark&\checkmark&\checkmark\\
 &$\rm A_{RF}$ &    $\rm m \slash s^2 $ & 0.3&&&&\checkmark&\checkmark&\checkmark&\checkmark\\
 &$\rm D_{RB}$ &    m & 1&&&&&&\checkmark&\checkmark\\
 &$\rm \rm V_{RB}$   & $\rm km \slash h $ & 1&&&&&&\checkmark&\checkmark\\
 &$\rm A_{RB}$ &    $\rm m \slash s^2 $& 0.3&&&&&&\checkmark&\checkmark\\
\midrule
 destination\tnote{d} &$\rm L_D$ &  - & -&&&&&&&\checkmark\\
 &$\rm D_D $ &  m & 1&&&&&&&\checkmark\\
\botrule\\
\end{tabular*}}{}
\begin{tablenotes}
\item[a] Each state is added with a noise from uniform distribution $\mathcal{U}$[-max noise,+max noise] before it is observed by car.
 \item[b] $\rm \delta$: front wheel angle;V: longitudinal velocity; A: longitudinal acceleration.
\item[c] $\rm L_C$: current lane (0 for left, 1 for right); $\rm V_{LU}$ and $\rm V_{LL}$: V - upper/lower speed limit of left lane, $\rm V_{RU}$ and $\rm V_{RL}$ are similar.
\item[d] $\rm L_D$: destination lane (0 for left 1, for right); $\rm D_D$: distance to the destination.
\end{tablenotes}
 \end{threeparttable}
\end{table*}

\subsubsection{Driving in lane}\label{sec:driving in lane}
For the driving-in-lane maneuver, the host car only needs to concern the traffic in its current lane from an ego-centric point of view. Therefore, we only need to design the state to model the current lanes. Furthermore, the SP-Net of this maneuver is only used for lateral control, and the state space for this situation does not need to include the information of the preceding car. Hence, as shown in Table \ref{tab.allstate}, the state space for the lateral control and longitudinal control are different. 

Intuitively, the reward for lateral control and longitudinal control should also be different. For lateral control learning, we use a reward function which provides a positive reward +1 at each step if the car drives in the lane, and a penalty of 0 for a terminal state of crossing the lane lines. For longitudinal control learning, we provide a positive reward proportional to the car's velocity if the speed is in the range of speed limit. The car would receive a penalty according to the extent of overspeed or underspeed, and a penalty of 0 for a terminal state of colliding with preceding vehicle. In summary, the reward function of driving-in-lane maneuver $r_{\text{d}}$ can be expressed as:
\begin{equation}
\nonumber
\begin{aligned}
&r_{\text{d,lateral}}=
\left\{
\begin{tabular}{ll}
0  & crossing lane\\
1 & else\\
\end{tabular}
\right.\\
&r_{\text{d,longit.}}=
\left\{
\begin{tabular}{ll}
2-0.01($\rm V_{\#U}-V$) & $\rm V_{\#L} \le V \le V_{\#U}$ \\
0.5+0.01($\rm V_{\#U}-V$) & $\rm  -20 \le V_{\#U}-V \le 0$ \\
0.5+0.01($\rm V-V_{\#L}$) & $\rm  -40 \le V-V_{\#L} \le 0$ \\
0  & else or collisions \\
\end{tabular}
\right.
\end{aligned}
\end{equation}
where the subscript $\# \in \{\text{R},\text{L}\}$ represents the right and left lanes.

\subsubsection{Right lane change}\label{sec:Right lane change}
For the right-lane-change maneuver, the car needs to concern the traffic in both the current lane and the right adjacent lane when making decisions. Therefore, we need to encode the information of the two lanes to define the state space for lane change maneuvers. Compared with the driving-in-lane maneuver, the lateral and longitudinal control in lane change share the same state (see Table \ref{tab.allstate}) and similar reward function. The sub-policy of lane change may contain both lane change and driving in lane functions. Such coupling of functions may lead to ambiguity in policy execution. Hence, a clear definition of lane change maneuver has been given to completely decouple the lane change maneuver and driving-in-lane maneuver. We designate the moment when the car center crosses the right lane line as the completion point of a right lane change. Note that completion does not necessarily mean success. If the car has crossed the right lane line, the learned sub-policy of Section \ref{sec:driving in lane} is then used to keep the car driving in the current lane. A lane change maneuver would be marked as success only if the host car does not collide with other cars or cross the lane lines in the next 5 s (200 steps). Otherwise, the lane change maneuver would be labeled as failure. 

For both the lateral and longitudinal control, we use a reward function assigning a penalty of -0.5 at each step and -100 for a terminal state of a failed lane change maneuver. Meanwhile, the car would receive a reward +100 for successful right lane change. In addition, the lateral control module would receive a penalty of -100 for a terminal state of crossing the left lane markings, and the longitudinal control module would receive a penalty of -100 for a terminal state of colliding with the preceding car. In summary, the reward function of right-lane-change maneuver $r_{\text{r}}$ can be expressed as:
\begin{equation}
\nonumber
\begin{aligned}
&r_{\text{r,lateral}}=
\left\{
\begin{tabular}{ll}
100 & success label\\
-100 & failure label\\
-0.5 & else
\end{tabular}
\right.\\
&r_{\text{r,longit.}}=
\left\{
\begin{tabular}{ll}
100 & success label\\
-100 & failure label or colliding before lane-change\\
-0.5 & else
\end{tabular}
\right.
\end{aligned}
\end{equation}
\subsubsection{Left lane change}
The state space and reward function of the left-lane-change maneuver is similar to Section \ref{sec:Right lane change}, except that the host vehicle needs to concern the left adjacent lane instead of right. The state space of this maneuver is also listed in Table \ref{tab.allstate} and the reward function is omitted. 
\subsubsection{Driving task} \label{sec:driving task}
Given the three existing maneuver policies, the car needs to concern all the states in Table \ref{tab.allstate} when making high-level maneuver selection. Firstly, we use a reward function to provide a positive reward of +1 at each step if the car drives in the high-speed lane and +0.5 for the low-speed lane. Of course, any maneuver leading to irregularities and collisions would receive a penalty of 0 along with a terminal signal. Unlike the learning of maneuver policies, master policy should consider the destination information. Hence, the car would receive a final reward of +200 if it reaches the destination. Otherwise, it would receive a penalty of 0. In summary, the reward function of the driving task $r_{\text{t}}$ can be expressed as:
\begin{equation}
\nonumber
r_{\text{t}}=
\left\{
\begin{tabular}{ll}
1 & high-speed lane\\
0.5 & low-speed lane\\
200 & reaching destination\\
0 & missing destination, irregularities, collisions\\
\end{tabular}
\right.
\end{equation}

Not all maneuvers are available in every state throughout driving. Considering the driving scenario of this study, the left-lane-change maneuver is only available in the right lane, and the right-lane-change maneuver is only available in left lane. Thus, we predesignate a list of available maneuvers via the observations at each step. Taking an unavailable maneuver is considered an error. The self-driving car should filter its maneuver choices to make sure that only legal maneuvers can be selected.

\section{Results} \label{sec:results}
The architecture of all the policy networks and value networks is the same except their output layers. For each network, the input layer is composed of the states followed by 2 fully-connected layers with 256 hidden units for each layer. We use exponential linear units (ELUs) for hidden layers, while the output layers of all the networks are fully-connected linear layers. The output of the value network is a scalar $V(s;w)$. On the other hand, the outputs of the policy network are a set of action preferences $h(a|s;\theta)$, one for each action. The action with the highest preference in each state is given the highest probability of being selected. Then, the numerical preferences are translated to softmax outputs which could be used to represent the selection probability of each action:
\begin{equation}
\nonumber
\pi(a|s;\theta) = \frac{{\rm exp}(h(a|s;\theta))}{\sum_a({\rm exp}(h(a|s;\theta)))}
\end{equation}

The optimization process runs 32 asynchronous parallel car-learners at a fix frequency of 40 Hz. During the training, we run backpropagation after $n = 10$ forward steps by explicitly computing 10-step returns. The algorithm uses a discount of $\gamma = 0.99$ and entropy regularization with a weight $\beta = 0.01$ for all the policies. The method performs updates after every 32 actions ($I_{update} = 32$) and shared RMSProp is used for optimization, with an RMSProp decay factor of 0.99. We learn the neural network parameters with a learning rate of $10^{-4}$ for all learning processes. For each episode, the maximum step of each car-learner is $t_{\text{max}}$ = 5000. 

\begin{figure*}[tb]
\centering
\captionsetup[subfigure]{justification=centering}
\subfloat[\textit{a}]{\label{subFig:pattern_0}\includegraphics[width=0.4\textwidth]{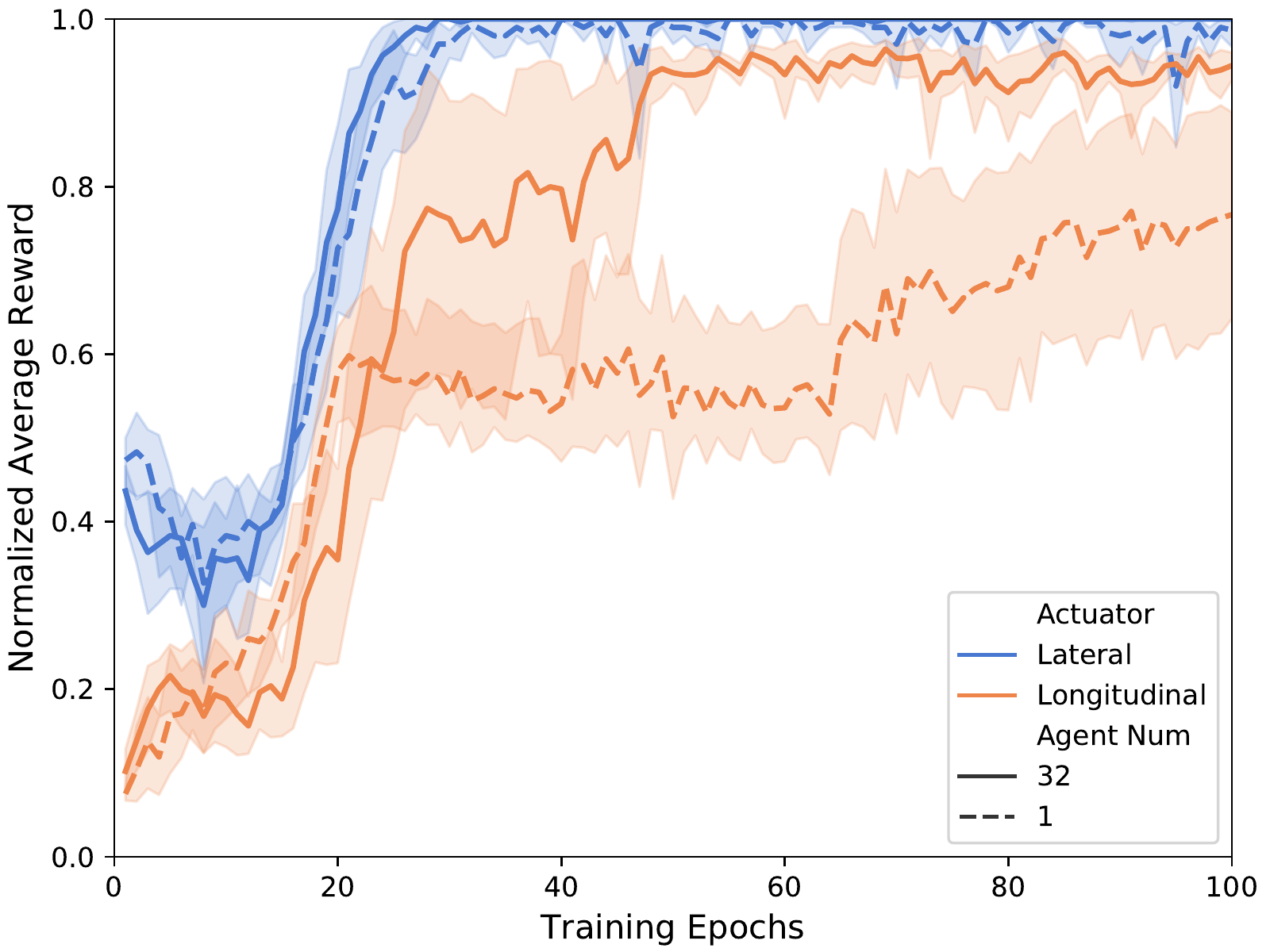}} 
    \subfloat[\textit{b}]{\label{subFig:pattern_2}\includegraphics[width=0.4\textwidth]{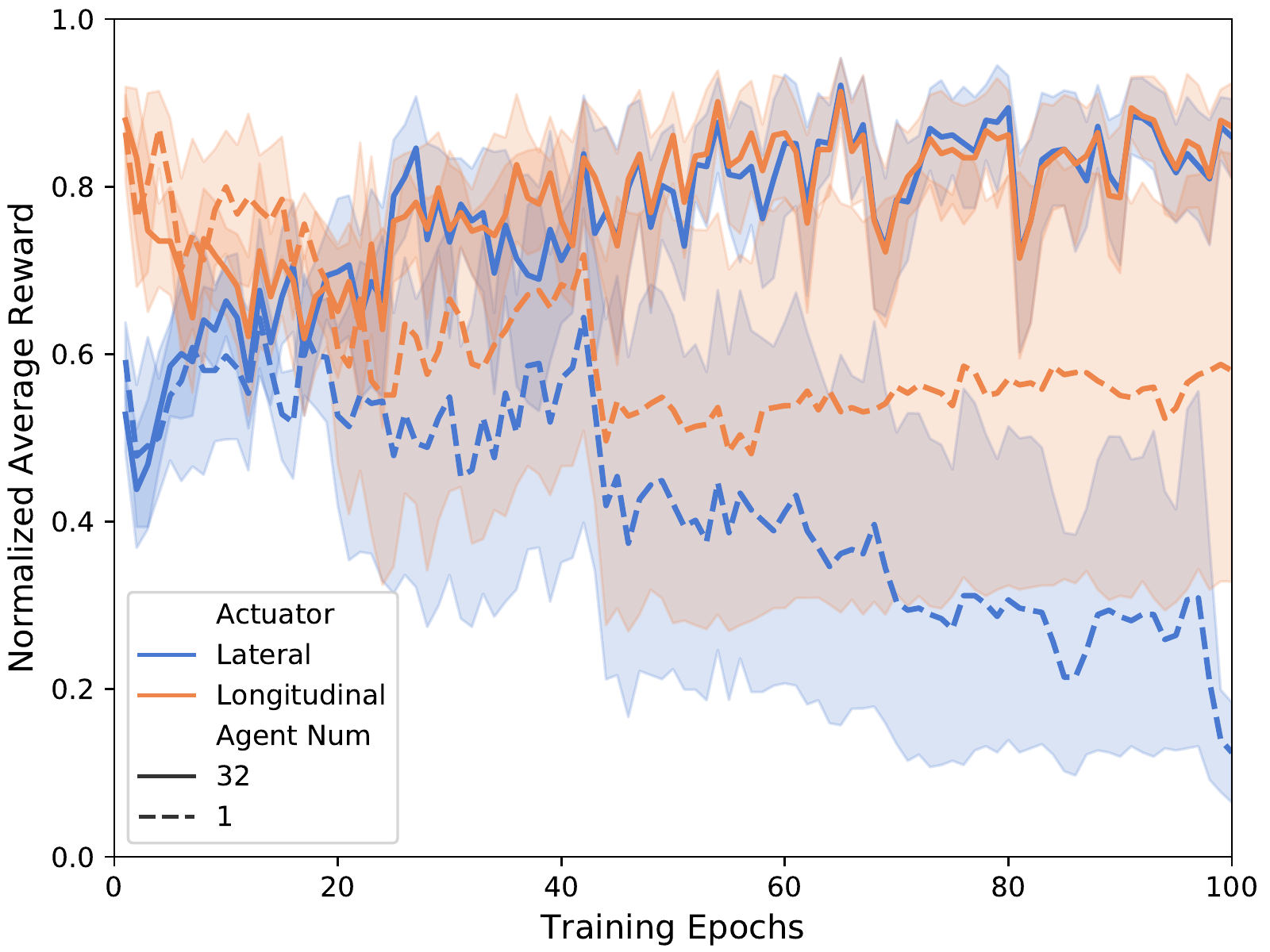}} \\
    \subfloat[\textit{c}]{\label{subFig:pattern_1}\includegraphics[width=0.4\textwidth]{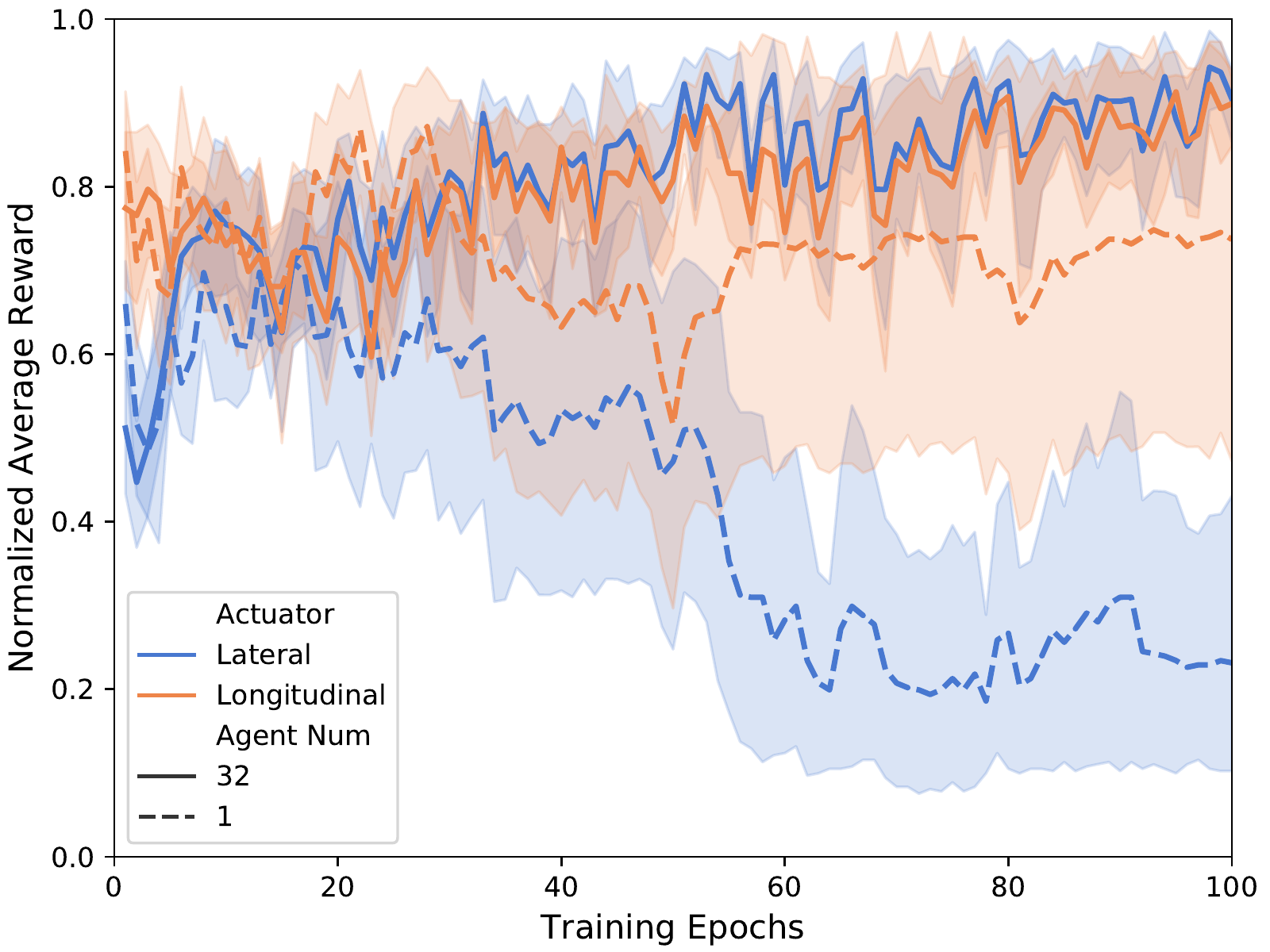}}
    \subfloat[\textit{d}]{\label{subFig:pattern_3}\includegraphics[width=0.4\textwidth]{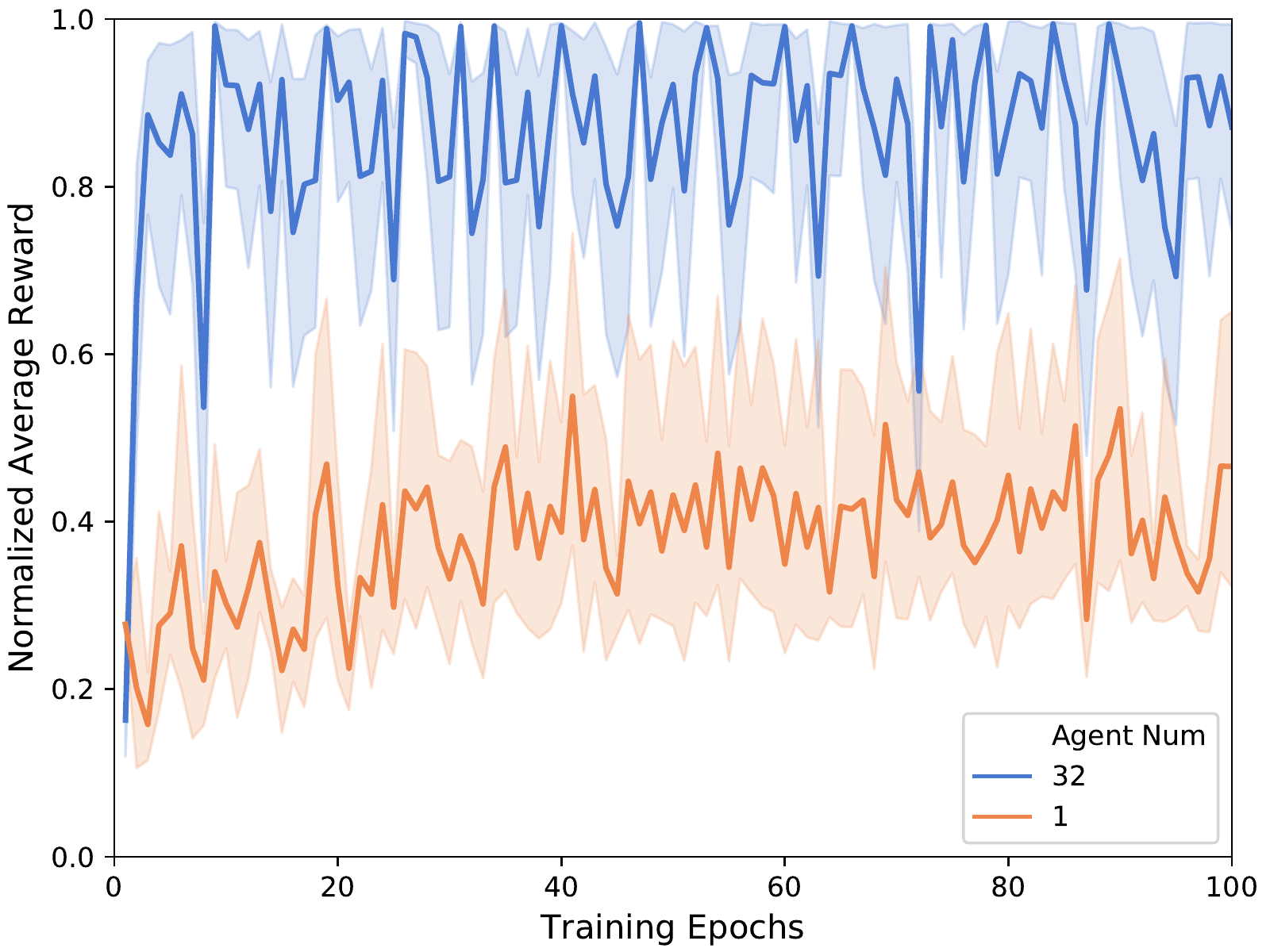}} \\ 
\captionsetup{font={footnotesize}}
\caption{{\normalsize \textit{Training performance (One epoch corresponds to 1000 updates for the shared networks)}}\\
\textbf{\textit{(a)}} Driving-in-lane maneuver, \textbf{\textit{(b)}} Right-lane-change maneuver, \textbf{\textit{(c)}} Left-lane-change maneuver, \textbf{\textit{(d)}} Maneuver selection}
\label{f:reward}
\end{figure*}
\subsection{Training performance}
We compare the training performance of the algorithm with 32 asynchronous parallel car-learners and that with only one car-learner. Fig. \ref{f:reward} shows the average value and 95\% confidence interval of normalized reward for 10  different  training  runs. The training results of driving-in-lane maneuver are shown in Fig. \ref{subFig:pattern_0}. As shown in this figure, both the learning speed, training stability and policy performance of APRL are much higher than the single-learner RL. On the other hand, it is easy to see that the SP-Net converges much faster than the AP-Net, which are nearly stable after approximately 30 epochs and 50 epochs, respectively. Besides, the lateral reward reaches a more stable plateau than the longitudinal reward. These noticeable differences are mainly caused by two reasons. First, the delay between the acceleration increment and resulting longitudinal rewards is too long. For example, the car usually needs to take hundreds of time steps to improve its speed to cause a rear-end collision. Nevertheless, the car may receive a penalty for crossing the road lane marking within only 20 steps. Second, the self-driving car is initialized with random velocity and position at the beginning of each episode. This would affect the longitudinal reward received by the car, because it is related to the velocity and speed limit. 

The training results of right-lane-change maneuver are shown in Fig. \ref{subFig:pattern_2}. For APRL algorithm, both the the lateral and longitudinal control modules stabilized after about 60 epochs. The fluctuation of the two curves around the peak after convergence is caused by the random initialization. The training results of left-lane-change maneuver are shown in Fig. \ref{subFig:pattern_1}, which are similar to results of the right-lane-change maneuver. On the other hand, the single-learner RL fails to learn suitable policies in these two cases because the state dimension is relatively high. Therefore, lane-change maneuver requires more exploration and is more complicated than driving-in-lane maneuver, which is too hard for single car-learner.

Fig. \ref{subFig:pattern_3} shows the training performance curve for the master policy. The results of APRL show that the master policy converges much more quickly than the maneuver policy, which reaches a plateau after around 10 epochs. The fluctuation is caused by the random initialization and the number of times the self-driving car reaches the destination in each epoch. In comparison, the learning speed and learning performance of  single-agent RL are very poor.

\begin{figure}[tb]
\centering
\captionsetup[subfigure]{justification=centering}
\subfloat[\textit{a}]{\label{subFig:decisionprogress}\includegraphics[width=0.45\textwidth]{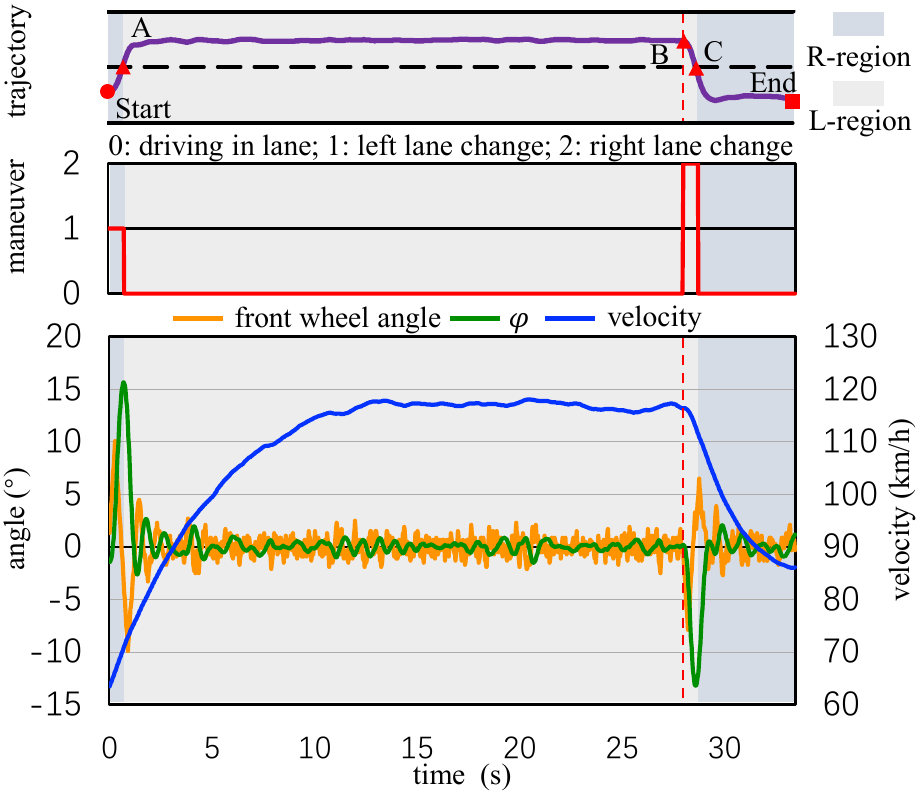}} \\
    \subfloat[\textit{b}]{\label{subFig:applicationtraffic}\includegraphics[width=0.45\textwidth]{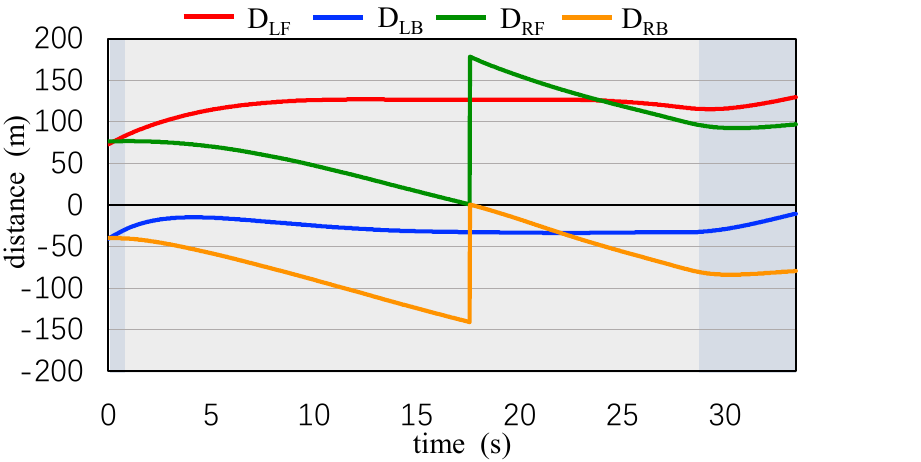}} \\ 
        \subfloat[\textit{c}]{\label{subFig:applicationtrafficV}\includegraphics[width=0.45\textwidth]{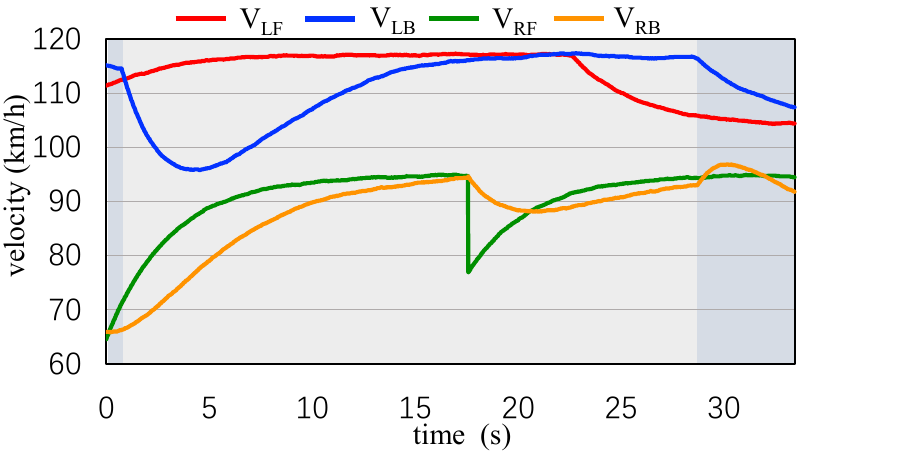}} \\ 
\captionsetup{font={footnotesize}}
\caption{{\normalsize \textit{Simulation results display }}\\
\textbf{\textit{(a)}} High-level maneuver selection and low-level motion control, \textbf{\textit{(b)}} Relative distance from other cars, \textbf{\textit{(c)}} Speed of other cars (R-region and L-region indicate that the ego vehicle is driving in the right lane and the left lane, respectively.)}
\label{f:simulation}
\end{figure}
\subsection{Simulation verification}\label{sec:simulation}

We conduct some simulations to assess the performance of the learned master policy and maneuver policies. The simulation platform used in this paper, including the traffic flow module and the vehicle model module, is directly built in Python. The states of all vehicles are randomly initialized for each simulation. The self-driving car is initialized on the low-speed lane with a random velocity. The destination is randomly placed between 800 and 1000 meters ahead of the car in the same lane. Random continuous traffic is presented in both lanes to imitate real traffic situations. Fig. \ref{subFig:decisionprogress} displays the decision making progress of both the master policy and the maneuver policies in a particular driving task. The position and velocity of other cars are shown in Fig. \ref{subFig:applicationtraffic} and \ref{subFig:applicationtrafficV}. Note that the sudden change in the curves in these two subgraphs (take the green line as an example) represents the change in relative position between the self-driving car and other cars. In particular, when the ego vehicle is driving in the left lane, $\text{D}_{\text{LF}}$  and $\text{V}_{\text{LF}}$ are the following distance and the speed of the preceding vehicle, respectively. Similarly, $\text{D}_{\text{RF}}$  and $\text{V}_{\text{LF}}$ have similar meanings when the ego vehicle is driving in the right lane. The car corresponding to this curve also changes at this time.

As shown in Fig. \ref{subFig:decisionprogress}, the self-driving car changes to the high-speed lane (from point Start to point A) at the beginning of the simulation and then keeps driving in this lane for a while. In the meantime, the velocity is increased in this phase from around 65 $\rm km \slash h $ to 118 $\rm km \slash h $. The self-driving car switches to low-speed lane (from point B to point C) when approaching the destination (point End), and the velocity eventually decreases to around 86 $\rm km \slash h $. Interestingly enough, the car tends to drive near the lane centerline based on its maneuver policy, but the reward function does not explicitly instruct it to do that. This is because that driving near the centerline has the highest state value with respect to the SV-Net of the driving-in-lane maneuver. Driving near the road lane marking will easily cause the lane departure due to the state noise and high speed. This demonstrates that our method is able to learn a rationale value network to evaluate the driving state. The master policy together with the three maneuver policies are able to realize smooth and safe decision making on highway.

\begin{figure}[b]
\centering
\captionsetup[subfigure]{justification=centering}
\subfloat[\textit{a}]{\label{subFig:resultcompare}\includegraphics[width=0.45\textwidth]{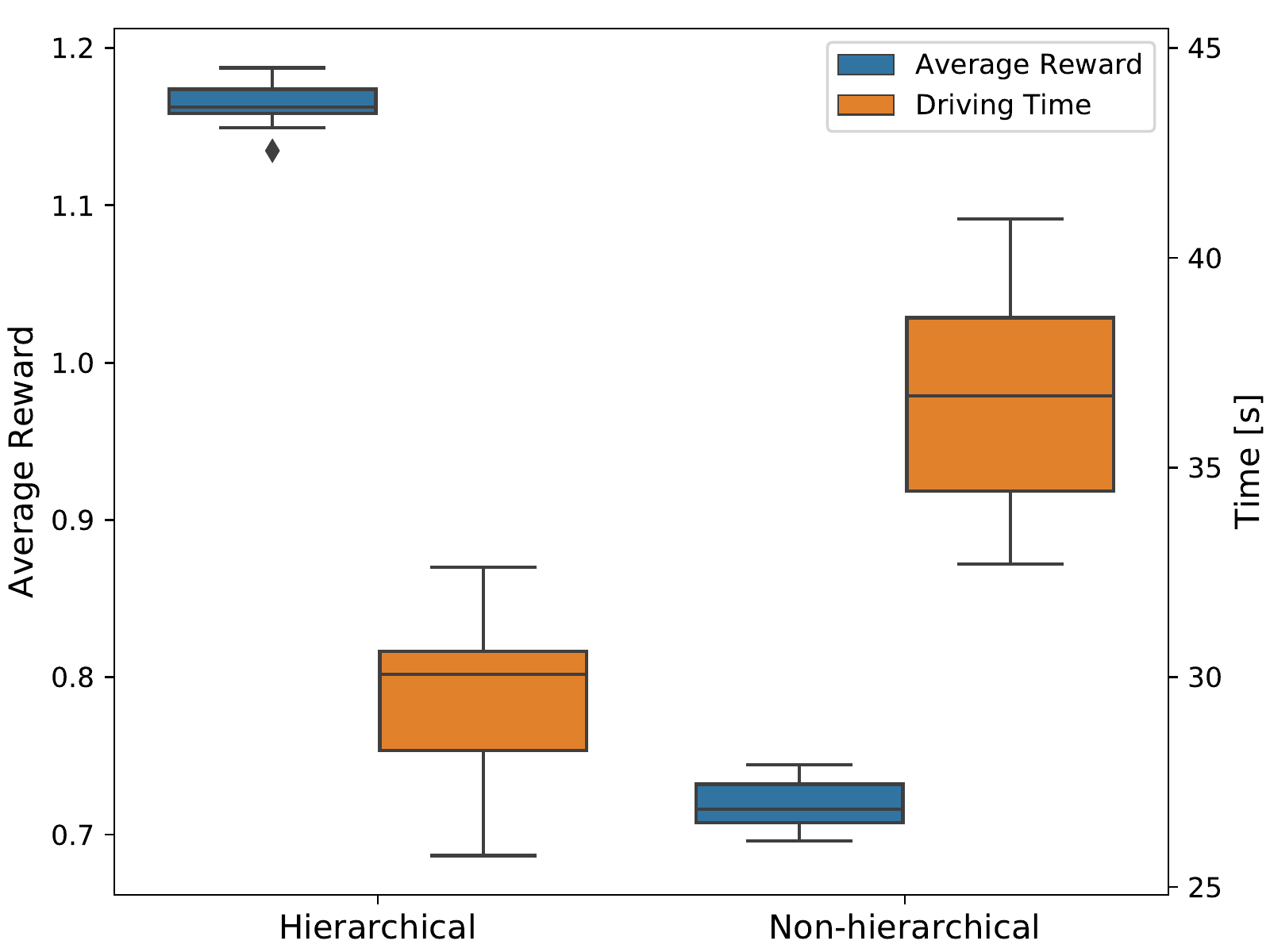}} \\
\subfloat[\textit{b}]{\label{subFig:trajectorycompare}\includegraphics[width=0.45\textwidth]{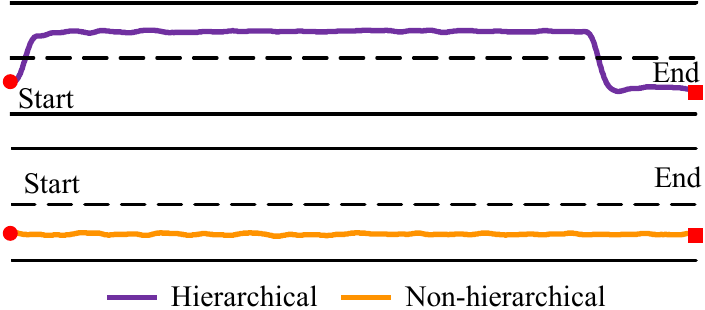}} \\ 
\captionsetup{font={footnotesize}}
\caption{{\normalsize \textit{Performance comparison}}\\
\textbf{\textit{(a)}} Average reward and time consumption per simulation, \textbf{\textit{(b)}} Driving trajectory}
\label{f:compare}
\end{figure}

We also directly trained non-hierarchical longitudinal and lateral driving policies using reward functions similar to those mentioned in Section \ref{sec:driving task}. Both networks take all the states in Table \ref{tab.allstate} as input, and the NN architecture is the same as the maneuver NNs mentioned above. Fig. \ref{subFig:resultcompare} shows the boxplots of the average reward and driving time for 20 different simulations of both hierarchical and non-hierarchical methods. Fig. \ref{subFig:trajectorycompare} plots typical decision trajectories for these two algorithms. It is clear that, policy of non-hierarchical method have not learned to increase the average driving speed by changing to the high-speed lane, resulting in lower average speed and longer driving time (about 25\% slower). Obviously, the policy found by the non-hierarchical method is local optimal solution. This is because during training process the self-driving often collides with other vehicle when changing to the other lane, so it thinks it is best to drive in the current lane. This is why we chose the hierarchical architecture for self-driving decision-making.

\subsection{Sensitivity to state noise}\label{sec:sensitivity}
As mentioned above, we add sensing noise to each state before being observed by the host car during training. Denoting the max noise of each state in Table \ref{tab.allstate} as $N_{\text{max}}$, we assume that the sensing noise obeys uniform distribution $\mathcal{U}(-N_{\text{max}},N_{\text{max}})$. In real application, the sensor accuracy is usually determined by sensor configuration and sensing algorithms \cite{de2017survey,jeng2013estimating,park2014robust}. Therefore, the sensor noise of certain states of a real car may be greater than $N_{\text{max}}$. We assume that the noise of all states in practical application is $M$ times the noise used in the training, so the real sensing noise obeys $\mathcal{U}(-M*N_{\text{max}},M*N_{\text{max}})$. To assess the sensitivity of the H-RL algorithm to state noise, we fix the parameters of previously trained master policy and maneuver policies and assess these policies for different $M$.

\begin{figure*}[!htb]
\centering
\captionsetup[subfigure]{justification=centering}
\subfloat[\textit{a}]{\label{subFig:noise_0}\includegraphics[width=0.4\textwidth]{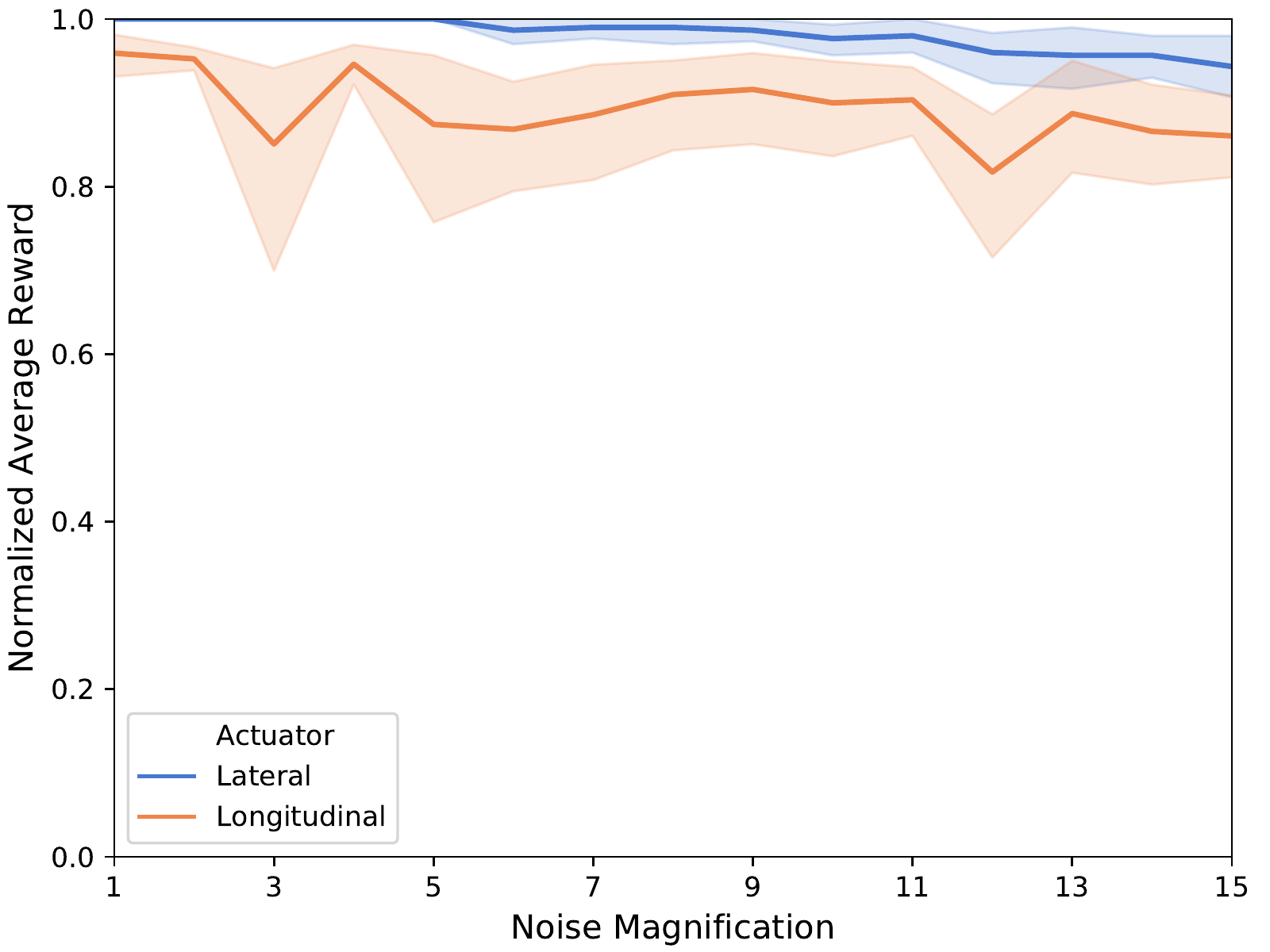}} 
    \subfloat[\textit{b}]{\label{subFig:noise_2}\includegraphics[width=0.4\textwidth]{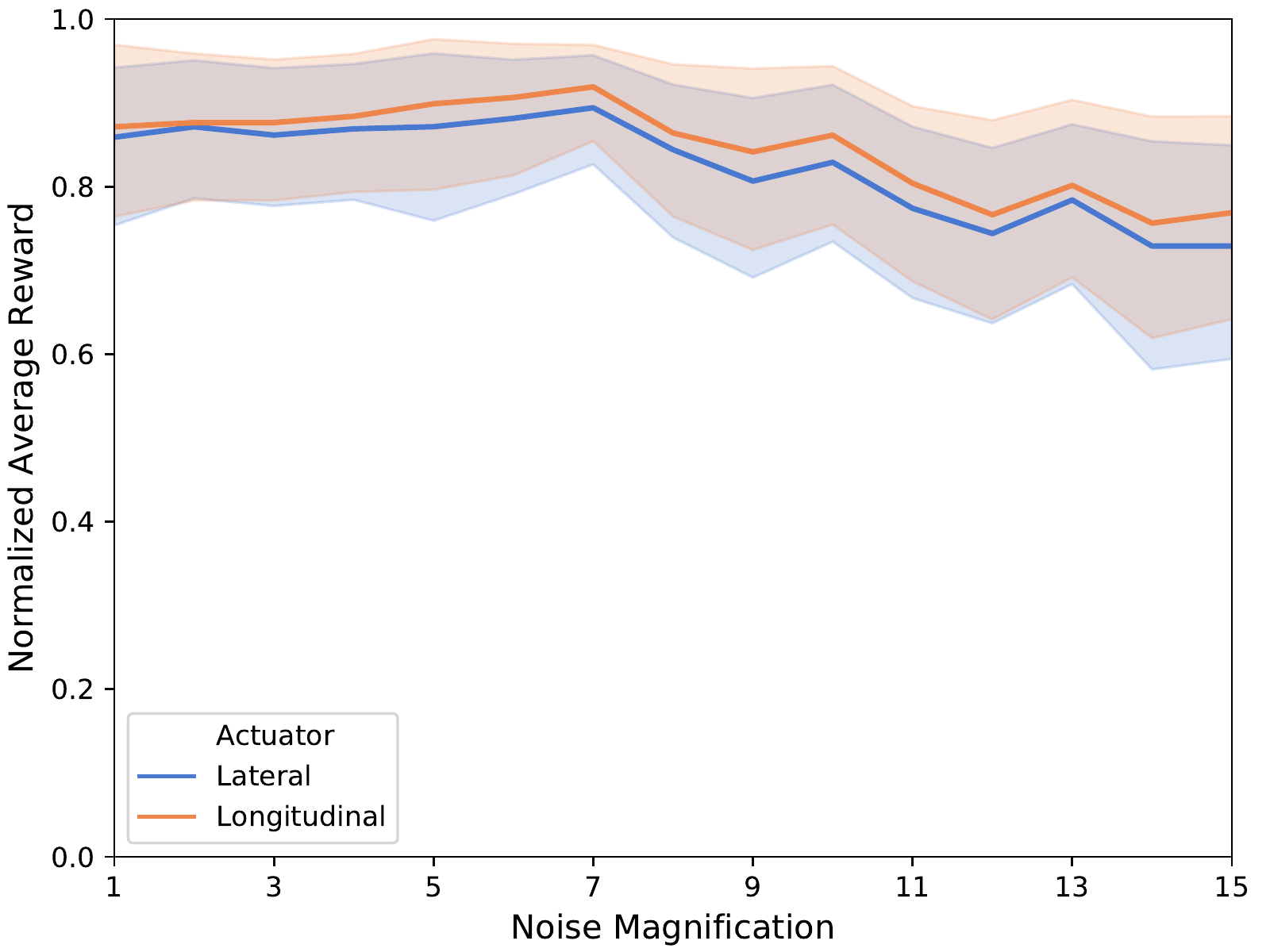}} \\
    \subfloat[\textit{c}]{\label{subFig:noise_1}\includegraphics[width=0.4\textwidth]{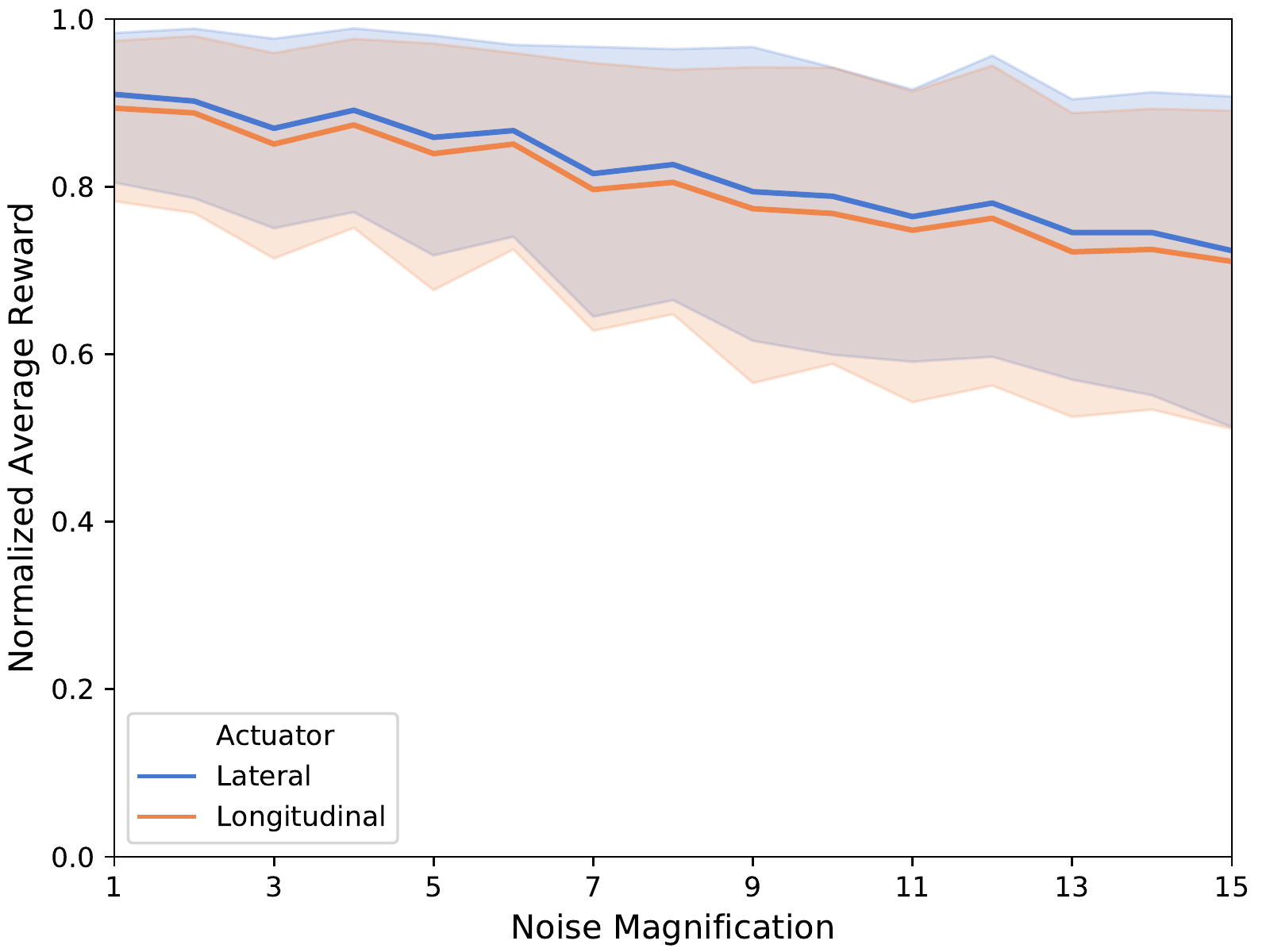}}
    \subfloat[\textit{d}]{\label{subFig:noise_3}\includegraphics[width=0.4\textwidth]{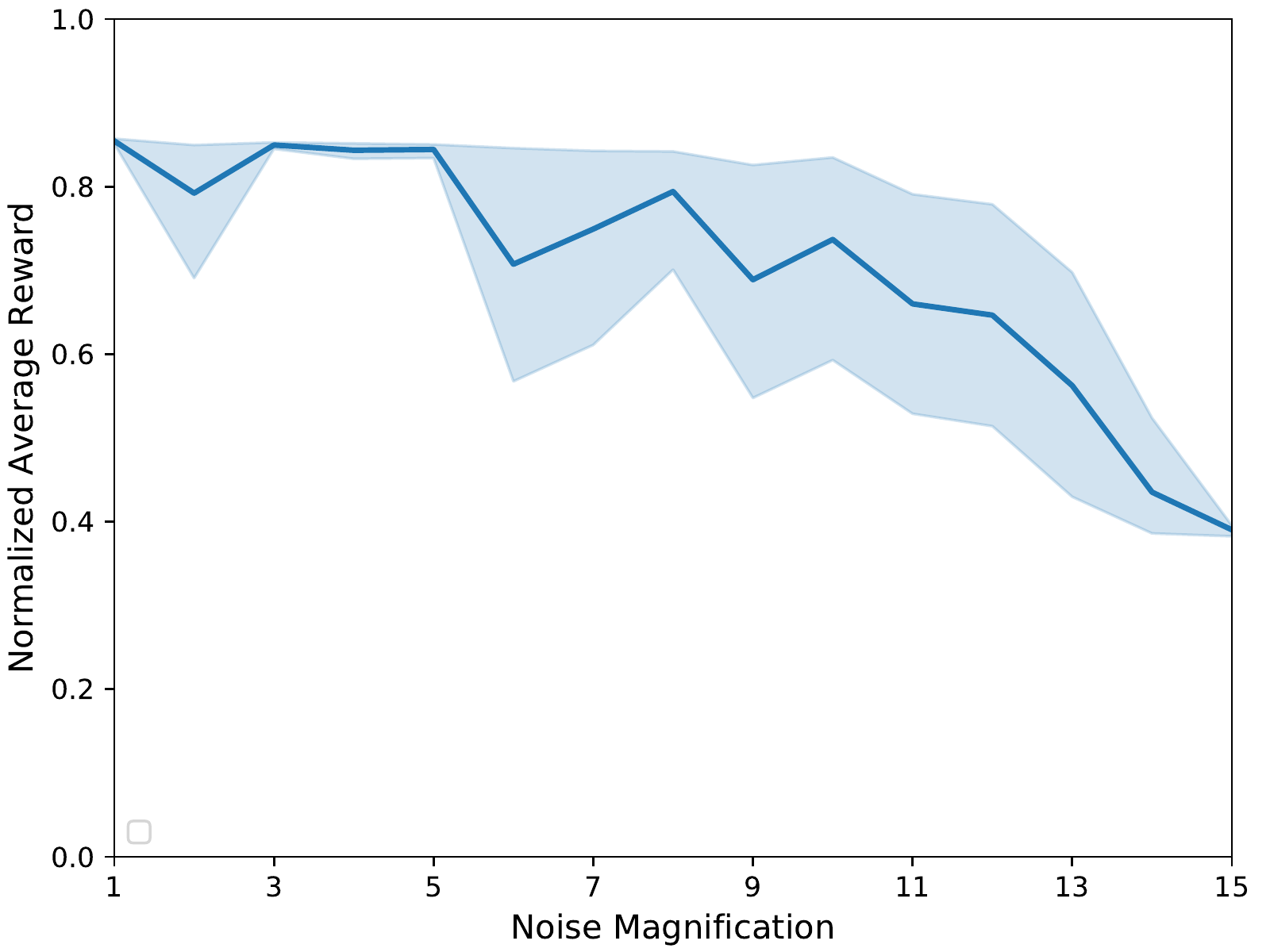}} \\ 
\captionsetup{font={footnotesize}}
\caption{{\normalsize \textit{Policy performance for different noise magnification $M$}}\\
\textbf{\textit{(a)}} Driving-in-lane maneuver, \textbf{\textit{(b)}} Right-lane-change maneuver, \textbf{\textit{(c)}} Left-lane-change maneuver, \textbf{\textit{(d)}} Maneuver selection}
\label{f:noise}
\end{figure*}

Fig. \ref{f:noise} shows the average value and 95\% confidence interval of normalized reward for 10 different trained policies. It is clear that the policy is less affected by noise when $M \le 7$. It is usually easy for the today's sensing technology to meet this requirement \cite{de2017survey,jeng2013estimating,park2014robust,guan2016use,cao2019geometry}. To a certain extent, this shows that the trained policies have the ability to transfer to real applications. Of course, it would be better if we add the same sensing noise as the real car during the training.

\section{Conclusion}\label{sec:conclusions}

In this paper, we provide a hierarchical RL method for decision making of self-driving cars, which does not rely on a large amount of labeled driving data. This method comprehensively considers the high-level maneuver selection and the low-level motion control in both the lateral and longitudinal directions. We first decompose the driving tasks into three maneuvers, i.e., driving in lane, right lane change and left lane change, and learn the sub-policy for each maneuver. The driving-in-lane maneuver is a combination of many behaviors including lane keeping, car following and free driving. Then, a master policy is learned to choose the maneuver policy to be executed at the current state. The  motion  of  the  vehicle  is  controlled  jointly  by the  lateral  and  longitudinal  actuators,  and  the  two  types  of actuators are relatively independent. All policies including master policy and maneuver policies are represented by fully-connected neural networks and trained using the proposed APRL algorithm, which builds a mapping from the sensory inputs to driving decisions. We apply this method to a highway driving scenario, and the state and reward function of each policy are designed separately, which demonstrates that it can realize smooth and safe decision making for self-driving cars. Compared with the non-hierarchical method, the driving time spent per simulation is reduced by approximately 25\%. In the future, we will continue to validate and improve our algorithms on highways with complex traffic, intersections and urban environment.

\section{Acknowledgments}\label{sec:ack}

This study is supported by International Sci\&Tech Cooperation Program of China under 2016YFE0102200 and NSF China with U1664263. We would also like to acknowledge Renjie Li and Long Xin for their valuable suggestions for this paper. 



\end{document}